\documentclass[a4paper,12pt]{article}
\usepackage{epsfig}
\usepackage{graphicx}
\usepackage{dcolumn}
\usepackage{bm}
\usepackage{longtable}

\usepackage{amssymb}
\usepackage{amsfonts}
\usepackage[margin=1.0in]{geometry}

\usepackage{amsmath}
\usepackage{multirow}
\allowdisplaybreaks[1]

\def\dd{\textrm{d}}  				     

\newcommand{\be}{\begin{equation}}
\newcommand{\ee}{\end{equation}}
\newcommand{\one}{{\rm 1\kern -.9mm l}} 

\usepackage{slashed}

\def\zetab{\boldsymbol{\zeta}}

\def\bB{\mathbf{B}}
\def\bF{\mathbf{F}}
\def\bH{\mathbf{H}}

\def\cD{ {\cal D} }

\def\ac{+\nonumber\\&}
\def\acI{+\right.\nonumber\\& \left.}
\def\acII{+\right.\right.\nonumber\\& \left.\left.}

\def\accI{\right.\nonumber\\& \left.}

\makeatletter							  
\@addtoreset{equation}{section}					  
\makeatother							  
\usepackage{hyperref}

\begin{document}
\begin{titlepage}
\begin{flushright}
DISIT-2013\\
DFPD-13/TH/03\\
CERN-PH-TH/2013-032
\par\end{flushright}
\vskip 2cm
\begin{center}
\textbf{\huge \bf Fermionic Corrections to Fluid}
\\\vspace{.5cm}
\textbf{\huge \bf Dynamics from BTZ Black Hole}
\textbf{\vspace{2cm}}\\
{\Large L.G.C.~Gentile$ ^{~a, c, e,}$\footnote{lgentile@pd.infn.it},$\ $ P.A.~Grassi$ ^{~a, d, f}$\footnote{pgrassi@mfn.unipmn.it}  and A.~Mezzalira$ ^{~b, d,}$\footnote{mezzalir@to.infn.it}}

\begin{center}
{a) { \it DISIT, Universit\`{a} del Piemonte Orientale,
}}\\
{{ \it via T. Michel, 11, Alessandria, 15120, Italy, }}
 \\ \vspace{.2cm}
 {b) { \it Dipartimento di Fisica Teorica, Universit\`a di Torino,}}\\
 {{\it via P. Giuria, 1, Torino, 10125, Italy,}}
\\ \vspace{.2cm}
{c) { \it Dipartimento di Fisica Galileo Galilei,\\
Universit\`a di Padova,\\
via Marzolo 8, 35131 Padova, Italy,
}}
 \\ \vspace{.2cm}
 {d) { \it INFN - Gruppo Collegato di Alessandria - Sezione di Torino,}}
 \\ \vspace{.2cm}
 {e) { \it INFN, Sezione di Padova,\\
via Marzolo 8, 35131, Padova, Italy.}}
 \\ \vspace{.2cm}
 {f) { \it
PH-TH Department, CERN,\\
CH-1211 Geneva 23, Switzerland.}}
\end{center}

\par\end{center}
\vfill{}

\begin{abstract}
{\vspace{.3cm}

       \noindent
We reconstruct the complete fermionic orbit of the non-extremal
BTZ black hole by acting with finite supersymmetry transformations.
The solution satisfies the exact supergravity equations of motion to
all orders in the fermonic expansion and the final result is given in terms
of fermionic bilinears. By fluid/gravity correspondence, we derive 
linearized Navier-Stokes equations and a set of new differential equations
from Rarita-Schwinger equation.
We compute the boundary energy-momentum tensor and we interpret the
result as a perfect fluid with a modified definition of fluid velocity. Finally, we
derive the modified expression for the entropy of the black hole in terms
of the fermionic bilinears.
}
\end{abstract}
\vfill{}
\vspace{1.5cm}
\end{titlepage}

\vfill
\eject

\tableofcontents
\setcounter{footnote}{0}


\section{Introduction}


Motivated by the success of fluid/gravity correspondence \cite{Bhattacharyya:2008jc,Rangamani:2009xk,Hubeny:2011hd},
we explore the connection between supergravity and a hypothetical supersymmetric fluid (we intend a fluid which is a long range
approximation of a supersymmetric theory) on the boundary of the $AdS$ space where a  black hole (BH) is located.
The reason for this analysis is rooted in the idea that by performing some perturbations around the black hole and promoting the
parameters of the infinitesimal isometry transformations to local parameters on the boundary, one is  able to derive
a set of partial differential equations for these parameters which can be identified with Navier-Stokes equations.
The fluid/gravity correspondence is obtained as follows: at first, one considers a solution of gravity equations such as
a black hole or a black brane (in our case a solution of supergravity with all fermionic zero modes), then one performs an isometry transformation of $AdS$ space
to obtain a new solution which, of course, will depend upon some constant parameters (such as the position or the scale).
Then, those parameters are promoted to fields of the boundary and the solution will no longer solve the equations of motion.
Nonetheless, one can see new partial differential equations emerging from that procedure which have an interpretation
as Navier-Stokes equations for the boundary fluid. For that, one needs to interpret the parameters of the isometries, namely
the translations or the scale, as the four velocity or the temperature of the fluid.

In some recent papers \cite{Gentile:2011jt,Grassi:2011wt}, we generalize that scheme to supergravity and to supersymmetric
fluids on the boundary. In particular, we have to recall that $AdS$ space is endowed with superisometries which introduce new
constant parameters in the solution. Again by promoting them to local fields on the boundary,
the solution will no longer solve the supergravity equations and new equations emerge
from imposing them. There are two problems to solve: 1) we have to start with a complete supergravity solution, namely we have to
take into account the full supermultiplet -- of which the black hole is the bosonic component -- in order to take  into account
the full orbit of the superisometries, 2) we have to promote the parameters of the superisometries to local fields on the boundary and
then interpret them as boundary fluid quantities.


In order to solve these problems, we adopt a simplifying framework where the computations can be done
analytically. We consider $\mathcal{N}=2$, $D=3$ supergravity with cosmological term
which has two solutions, the $AdS_3$ space and the BTZ black hole \cite{Banados:1992wn, Banados:1992gq}.
In a previous paper \cite{Gentile:2012tu}, we computed the full supermultiplet of the BTZ black hole
by performing a finite susy transformation. Notice that only by finite susy transformations, we are able to
compute the complete orbit ({\it wig}) starting from the black hole solutions \cite{Aichelburg,Gentile:2012jm}.
That susy transformation truncates at the forth order.
In the present paper, we parametrize the order of
computation by the powers of bilinears in Grassmann parameters.


In order to generate the complete wig we start
from the susy transformations associated to the Killing spinors of $AdS$ space. Since the BTZ black hole
is non-extremal any transformation will produce a change in the solution. Multiple applications of the susy transformations
generated by Killing spinors will result in the application of the corresponding Killing vector generating the complete
supergroup of isometries of $AdS$ space which is ${OSp}(2|2)\times{OSp(2|2)}/SO(2)\times OSp(2|2)$.\footnote{$AdS$ space considered here is actually a superspace
with 3 bosonic coordinates and 4 fermionic coordinates, it can be viewed as $OSp(2|2)/SO(2)$ (since $Sp(2)  \sim SL(2,\mathbb R)
\sim SO(1,2)$).}


Given the new solution, one can observe that the some isometries of the black hole such as the translation invariance
in the time direction and in the angular coordinate (or in the space coordinate in the Poincar\'e patch) are preserved. That implies that
the mass $M$ and the angular momentum $J$ are still conserved charges. Indeed, we can compute them using the ADM formalism and
that gives a mass and an angular momentum which is shifted by fermionic bilinears. In the case of extremal black hole, $M = |J|$, the fermionic
corrections will not spoil the extremality condition. In the same way, we can compute also
the entropy of the black hole which is modified by fermionic bilinears.


Having set up the stage for the computation, we promote the fermionic parameters of the superisometries to
local parameters on the boundary. Then, by inserting the fields in the supergravity equations we find
two sets of new equations which should be satisfied: Navier-Stokes equations (which we also computed in
previous paper \cite{Gentile:2011jt}) and new differential equations for the fermionic degrees of freedom.
In order to interpret the result obtained we also perform the bosonic isometries associated to the dilatation
and to the translation on the boundary reproducing the usual linearized
version of relativistic Navier-Stokes equations. On the other side, by inserting the
solution in the gravitino equation, we finally derive a new set of partial differential equations for
the fermionic degrees of freedom which we interpret as Dirac-type equation for the fluid.


With the complete metric, we can finally compute the extrinsic curvature and, using
Brown-York  procedure \cite{Brown:1992br,Balasubramanian:1999re} we derive the boundary
energy-momentum tensor. The form of the latter resemble the tensor for a perfect fluid, except
for a term (violating the chirality). Nonetheless a redefinition of the velocity of the fluid takes the
energy momentum tensor to the standard formula for a perfect fluid and the temperature is shifted by terms dependent on bilinears.
The computation has been performed at the first level in the isometry parameters and it shows the absence of dissipative effects, as expected from a conformal fluid in $1+1$ dimensions.
To see the emergence of new structures in the fluid energy--momentum tensor one needs a complete second order computation.


In sec. 1, we set up the stage for the computation. In sec. 2 we present the complete wig solution
of the black hole. We also provide the expressions for large $r$ which are useful for checking the
structure of the solution. In sec. 3, we derive the new differential equations on the boundary of $AdS$ and we
compute the Dirac-type equation on the boundary. In sec. 4, we compute the energy-momentum tensor and
we discuss the redefinition of the fluid velocity to reabsorb the parity--violating term.


\section{Setup}

\subsection{Action and Equations of Motion}

As mentioned in the introduction, we consider the supersymmetric $\mathcal{N}=2$, $D=3$ of \cite{Achucarro:1987vz,Izquierdo:1994jz} whose field content is described by the vielbein $e^{A}$, the gravitino (complex) $\psi$, an abelian gauge field $A$ and the spin connection $\omega^{AB}$.
Those are the gauge fields of the diffeomorphism, the local supersymmetry, the local $U\left( 1 \right)$ transformations and of the Lorentz symmetry.
The gauge symmetry can be used to gauge out all local degrees of freedom and the remaining d.o.f. are localized singular solutions \cite{Banados:1992wn,Banados:1992gq,Henneaux:1999ib,Henneaux:1984ei}.

The invariant action has the following form
\begin{align}
	S = & \int_{\mathcal{M}} \left(
		R^{AB} \wedge e^{C} \varepsilon_{ABC}	
		-
		\frac{\Lambda}{3} e^{A} \wedge e^{B} \wedge e^{C} \varepsilon_{ABC}
		-
		\bar \psi \wedge {\cal D} \psi
		-
		2 A \wedge \dd A
		\right)
		\ ,
	\label{action0}
\end{align}
where the curvature $2$--form is defined as $R^{AB}= \dd \omega^{AB} + \omega^{A}{}_{C}\wedge \omega^{CB}$ and $\mathcal{M}$ is a $3d$ manifold.
In components, the action reads
\begin{align}
	S= \int\! d^3x\, \left[
		e \left( R + 2 \Lambda \right)
		- \bar\psi_{M} {\cal D}_{N} \psi_{R} \varepsilon^{MNR}
		- 2 A_{M} \partial_{N} A_{R} \varepsilon^{MNR}
	\right]\ ,
\label{action1}
\end{align}
where $e$ is the vielbein determinant and $R$ is the Ricci scalar\footnote{$\left\{ A,B,\dots \right\}$ label flat indices and $\left\{ M,N,\dots \right\}$ refer to curved ones}.
The action, is invariant under all gauge transformations and it can be cast in a Chern--Simons form\cite{Achucarro:1987vz}\footnote{Note that $AdS_{3}$ radius is set to one and $\left(8 \pi G_3\right)^{-1} = 1$.}.
The covariant derivative ${\cal D}_M$ is defined as
\begin{align}
	{\cal D}_M = D_M + i A_M - \frac{\Lambda}{2}e_{\phantom{A}M}^A \Gamma_A \ ,
\label{action2}
\end{align}
where $D= \dd + \frac{1}{4}\omega^{AB} \Gamma_{AB}$ is the usual Lorentz-covariant differential.
It can be easily shown that (\ref{action1}) is invariant under the local supersymmetry transformations
\begin{align}
\delta_\epsilon\psi &= {\cal D}\epsilon \ ,&
\delta_\epsilon e^A &= {\frac14}\left( \bar\epsilon\Gamma^A\psi - \bar\psi\Gamma^A\epsilon \right)
\ ,&
\delta_\epsilon A &= {\frac{i}{4}} \left( \bar\epsilon \psi - \bar\psi\epsilon \right) \ .
\label{action4}
\end{align}
The spin connection transforms accordingly when the vielbein postulate is used to compute $\omega^{AB}$.
The signature for the flat metric $\eta_{AB}$ is
$(-,+,+)$ and the gamma
matrices $\Gamma^A$ are real
\begin{align}
	\Gamma_{0} = &\  i \sigma_{2}
	\ , &
	\Gamma_{1} = &\  \sigma_{3}
	\ , &
	\Gamma_{2} = &\  \sigma_{1}
	\ , &
	\left\{ \Gamma_{A}\,,\,\Gamma_{B} \right\} = & 2 \eta_{AB}
	\ .
	\label{gammaBTZ}
\end{align}
From (\ref{action1}) we deduce the following equations of motion
\begin{align}
	& \cD \psi = 0
	\ , \nonumber \\
	& \dd A = \frac{i}{4} \bar \psi \wedge \psi
	\ , \nonumber \\
	& \dd e^{A} + \omega^{A}_{\phantom{A}B} \wedge e^B = {\frac14}\bar\psi\wedge\Gamma^A \psi
	\ ,\nonumber \\	
	& d \omega^{AB} + \omega^{A}{}_{C}\wedge \omega^{CB} - \Lambda e^{A} \wedge e^{B} = - \frac{\Lambda}{4} \varepsilon^{AB}{}_{C} \bar\psi\Gamma^{C}\psi
	\ .
	\label{EoMs1}
\end{align}
The third equation is the vielbein postulate, from which the spin connection $\omega^{AB}$ is computed.
It is possible to check the above equations against the Bianchi identities.
Note that the theory, being topological, can be written in the form language.\footnote{
Using the forms, the gauge symmetries are obtained by shifting all fields
 $e^{A}\rightarrow e^A + \xi^{A}$, $\psi\rightarrow\psi+\eta$, $\omega\rightarrow\omega^{AB} + k^{AB}$ and $A\rightarrow A+C$
and consequently the differential operator $\dd\rightarrow \dd+s$. $\xi^{A}$, $\eta$, $k^{AB}$ and $C$ are the ghosts associated to diffeomorphism, supersymmetry, Lorentz symmetry and $U\left( 1 \right)$ transformation, respectively and $s$ is the BRST differential associated to those gauge symmetries.}
The gravitino equation simply implies the vanishing of its field strength.
The second equation fixes the field strength of the gauge field and the fourth one fixes the Riemann tensor.
Note that for $AdS_{3}$, the cosmological constant is $\Lambda=-1$.

\subsection{$AdS_3$ and BTZ Black Hole}

The supergravity equations of motion admit as solution the $AdS_{3}$ space
\begin{align}
	& g_{MN} = (g_{AdS})_{MN} \ , &
	& A_{M} = 0 \ ,&
	\psi_{M} = 0 \ ,	
	\label{emptyAdS}
\end{align}
where the $AdS_{3}$ metric in global coordinates reads
\begin{align}
	\dd s^{2} & =
	-(1+r^2) \dd t^{2}
	+
	\frac{1}{1+r^2} \dd r^{2}
	+
	r^{2} \dd \phi^{2}
	\ ,
	\label{AdS}
\end{align}
Another solution is the so called BTZ black hole\footnote{We refer the reader to the vast literature on the subject for the geometry of this solution.} whose global metric reads:
\begin{align}
	\dd s^{2} =
	-N^{2} \dd t^{2}
	+
	N^{-2} \dd r^{2}
	+
	r^{2} \left( N^{\phi}\dd t + \dd \phi \right)^{2}
	\ ,
	\label{BTZ1}
\end{align}
where $N$ and $N^{\phi}$ are defined as
\begin{align}
	N = & \sqrt{- M_{0} + r^{2} + \frac{J_{0}^{2}}{4 r^{2}}}
	\ ,
	& N^{\phi} = & - \frac{J_{0}}{ 2 r^{2}}
	\ .
	\label{BTZ2}
\end{align}
The non--zero vielbein components  are
\begin{align}
	e^{0} = & N \dd t \ ,
	&
	e^{1} = & N^{-1} \dd r
	\ , &
	e^{2} = & r N^{\phi} \dd t + r \dd \phi
	\ ,
	\label{BTZviel}
\end{align}
and the non--zero spin connection components read
\begin{align}
	\omega^{0}{}_{1}= &
	r \dd t - \frac{J_{0}}{2 r} \dd \phi
	\ , &
	\omega^{0}{}_{2} = & - \frac{J_{0}}{2 r^{2} N} \dd r
	\ , &
	\omega^{1}{}_{2} = & - N \dd \phi
	\ .
	\label{BTZspinC}
\end{align}
The parameter $M_{0}$ is to be identified with the mass of the black hole while $J_{0}$ represents its angular momentum.
Setting $M_0 = -1$ and $J_0 = 0$ in (\ref{BTZ1}) we obtain the $AdS$ metric in global coordinates (\ref{AdS}).\footnote{Note that the region $-1<M_0<0$ is excluded since it corresponds to a naked singularity.}

To analyze the boundary fluid dynamic using fluid/gravity technique is convenient to consider $AdS_{3}$ metric written in a Poincar\'e patch
\begin{align}
	\dd s^{2} & =
	- r^2 \dd t^{2}
	+
	\frac{1}{r^2} \dd r^{2}
	+
	r^{2} \dd x^{2}
	\ .
	\label{AdSPP}
\end{align}
As in \cite{Bhattacharyya:2008jc,Rangamani:2009xk} we perform an ultralocal analysis and then we also use a Poincar\'e patch for BTZ black hole metric
\begin{align}
	\dd s^{2} & =
	-\left( r^{2} - M_{0} \right) \dd t^{2}
	+
	\frac{1}{r^2 - M_{0}} \dd r^{2}
	+
	r^{2} \dd x^{2}
	\ .
	\label{BTZPP}
\end{align}
Notice that in this case  the $AdS_{3}$ metric (\ref{AdSPP}) is obtained by setting $M_{0}=0$. The form of the metric is similar to (\ref{BTZ1}) but it will cover just a sector of the entire $AdS$ space.
As we will show in the next section, after a finite boost transformation the metric (\ref{BTZPP}) can be cast as in (\ref{BTZ1}), with mass and angular momentum depending on the boost parameters and $M_{0}$.

\subsection{Killing Vectors and Killing Spinors}

In this section we compute the Killing vectors and the Killing spinors for $AdS_{3}$. As we will discuss later, we consider the isometries of $AdS_3$ space to generate orbits of the black hole solution. This is obtained by acting with the generators of $AdS_3$ isometries on the black hole metric.

The Killing vectors are solutions to the equations
\begin{equation}
	\mathcal{L}_{\xi} (g_{AdS}) =  0
	\ ,
\end{equation}
where
\begin{equation}
	\xi = \xi^t \partial_t + \xi^r \partial_r + \xi^{\phi} \partial_{\phi}
	\ ,
\end{equation}
and, for global $AdS_{3}$ (\ref{AdS}), they are
\begin{align}
\xi^t & = \frac{r}{\sqrt{1+r^2}} \partial_t A\left(t,\phi\right) + e_0 \ ,
\nonumber\\
\xi^r & = \sqrt{1+r^2} A\left(t,\phi\right) \ , \nonumber
 \\
\xi^\phi & = \frac{\sqrt{1+r^2}}{r} \partial_{\phi} A\left(t,\phi\right) + f_0 \ ,
\label{KVglobal}
\end{align}
where the function $A\left(t,\phi\right)$ is defined as
\begin{equation}
A(t,\phi) = a_0 \cos\left(t+\phi\right)+ b_0 \cos\left(t-\phi\right) + c_0 \sin\left(t+\phi\right)+ d_0 \sin\left(t-\phi\right) \ .
\end{equation}
The solution depends upon the $6$ free parameters $\{a_{0},b_{0},c_{0},d_{0},e_{0},f_{0}\}$, associated to the $AdS_3$-isometry group, namely, $SO\left(2,2\right)$.

The Killing vectors for $AdS_{3}$ in Poincar\'e patch, defined as $K=K^t \partial_t + K^r \partial_r + K^x \partial_x$ are
\begin{align}
	K^{t}
	& =
	- \frac{c_{1}}{2}
	\left( \frac{1}{r^{2}} +t^{2} + x^{2} \right)
	-
	c_{2} t x
	-
	b t + w x + t_{0}
	\ , \nonumber \\
	K^{r}
	& =
	r \left(
	c_{1} t + c_{2} x + b
	\right)
	\ , \nonumber \\
	K^{x}
	& =
	- \frac{c_{2}}{2}
	\left( - \frac{1}{r^{2}} +t^{2} + x^{2} \right)
	-
	c_{1} t x
	+
	w t - b x + x_{0}
	\ .
	\label{KVPP}
\end{align}
The $6$ real infinitesimal constant parameters describe the $6$--parameters isometry group of $AdS_{3}$: $b$ is associated with dilatation, $w$ is the boost parameter, $c_{1}$ and $c_{2}$ are related to conformal transformations and $t_{0}$ and $x_{0}$ parameterize the $t-$ and $x-$translations.

In order to complete the procedure outlined in \cite{Bhattacharyya:2008jc,Rangamani:2009xk,Gentile:2011jt,Gentile:2012tu} we perform a finite boost on the
BTZ solution in the $t$-$x$ plane, namely
\begin{align}
	& t \rightarrow \frac{t - w x}{\sqrt{1 - w^{2}}}
	\ ,
	&
	x \rightarrow \frac{x - w t}{\sqrt{1 - w^{2}}}
	\ ,
	&
	\label{boost1}
\end{align}
where $w$ is the boost parameter.

We now perform a finite dilatation of the BTZ black hole. This transformation will allow us to define a parameter for the temperature of the fluid in the same fashion as \cite{Bhattacharyya:2008jc}.
The correct dilatation weights can be obtained by redefining the coordinates as follows
\begin{equation}
\label{dilatate}
r \rightarrow \hat b r
\ , \qquad \qquad
t \rightarrow \hat b^{-1} t
\ , \qquad \qquad
\phi \rightarrow \hat b^{-1} \phi
\ .
\end{equation}
The infinitesimal dilatation are given by
$\hat b = 1 + b + O(b^{2})$,
where $b$ is the infinitesimal parameter introduced in eq.~(\ref{KVPP}).

The boosted and dilatated metric can be recast in the form (\ref{BTZ1}) by replacing
\begin{align}
	&
	M_{0} \rightarrow M = \frac{1 + w^{2}}{1 - w^{2}} \frac{M_{0}}{\hat b^{2}}
	\ ,
	&
	J_{0} \rightarrow J =
	- \frac{2 w}{1 - w^{2}}\frac{M_{0}}{\hat b^{2}}
	\ ,
	&
	\label{boost2}
\end{align}
and
\begin{align}
	r^{2} \rightarrow R^{2}
	=
	r^{2}
	+
	\frac{w^{2}}{1-w^{2}} \frac{M_{0}}{\hat{b}^2}
	\ .
	\label{boost3}
\end{align}
Note that the boost transformations can be applied to the global BTZ metric (\ref{BTZ1}) to generate a new set of solutions, as described in \cite{Martinez:1999qi} (see eq.~(\ref{boost1}) with $x$ substituted by the angular coordinate $\phi$). In this case the replacing rules for mass, angular momentum and radius coordinate will be
\begin{subequations}
\begin{align}
&
	M_{0} \rightarrow M = \frac{1 + w^{2}}{1 - w^{2}} M_{0} - \frac{2 w}{1 - w^{2}} J_{0} \ , \\
	&
	J_{0} \rightarrow J = \frac{1 + w^{2}}{1 - w^{2}} J_{0} - \frac{2 w}{1 - w^{2}} M_{0} \ , \\
	&
	r^{2} \rightarrow R^{2}
	=
	r^{2}
	-
	\frac{w}{1-w^{2}} \left( J_{0} - w M_{0}  \right)
	\ .
\end{align}
\end{subequations}
Defining
\begin{align}
	&
	\gamma = \frac{w^{2} + 1}{w^{2} - 1}
	\ ,
	&
	\beta = - \frac{2 w}{ w^{2} + 1}
	\ ,
	&
	\label{boost4}
\end{align}
the metric for the new global BTZ solutions can be written modifying mass and angular momentum in the following Lorentz-like form, \textit{i.e.}
\begin{align}
M& = \gamma M_0 - \beta \gamma J_0 \ , \nonumber\\
J& = \gamma J_0 - \beta \gamma M_0 \ , \nonumber\\
R^2& = r^2 -\frac{1}{2} \left[\beta J_0 - \left(\gamma +1\right)M_0\right]
\ .
\end{align}

Now, we need to construct the Killing spinors of $AdS_3$ and the isometries generated by them\footnote{We remind the reader that Killing vectors can be obtained constructing Killing spinors bilinears such as $\xi^{\mu}= \bar{\epsilon}\Gamma^{\mu}\epsilon$. By construction they will indeed satisfy the Killing vectors equation.}.
To construct the BTZ wig we compute the Killing spinors $\epsilon $ for $AdS_{3}$ Poincar\'e patch, defined from  Killing spinors equation
\begin{align}
	\mathcal{D}_{AdS} \epsilon = 0 \ .
	\label{KSeq}
\end{align}
We have
\begin{equation}
	\epsilon
	 =
	\left[
	\frac{1}{2\sqrt{r}} \left( \one - r x^{\mu} \Gamma_{\mu}\right)
	\left( \one +\Gamma_{1} \right)
	+\frac{\sqrt{r}}{2} \left( \one - \Gamma_{1} \right)
	\right]
	\zeta\ ,
	\label{ciao}
\end{equation}
where $\zeta$ is a Dirac spinor with $2$ complex constant components $\zeta_{1}$ and $\zeta_{2}$
\begin{align}
	\partial_{R} \zeta = 0
	\ .
	\label{constantZeta}
\end{align}
See also \cite{Gentile:2012tu} where a deeper analysis of the Killing spinor is performed.


\section{Fermionic Wig}

We now proceed to the construction of the fermionic wig (\textit{i.e.} the complete solution in the fermionic zero modes) associated with a boosted and dilatate BTZ black hole in Poincar\'e patch.\footnote{Note that our Killing spinors (or anti-Killing spinors as defined in \cite{Kallosh:1996qi}) are not time independent but the fermionic black hole superpartner does not depend on $t$.}
As explained in the previous section, the boost and the dilatation shift the mass and angular momentum of the black hole. Therefore, to get the complete solution, we
first compute the wig for the BH and then we perform the shift to the mass and of the angular momentum.

Thus, we proceed in the usual way by constructing the wig for the black hole metric (\ref{BTZ1}) and then replacing $M_0$ and $J_0$ with $M$ and $J$ as defined in (\ref{boost2}).
This procedure is iterative and can be found in \cite{Gentile:2012tu}.
The superpartner of a generic field $\Phi$ is constructed by acting with a finite supersymmetry transformation on the original field \cite{Burrington:2004hf}:
\begin{align}
	\boldsymbol{\Phi} = e^{\delta_{\epsilon}} \Phi = \Phi + \delta_{\epsilon} \Phi + \frac{1}{2} \delta_{\epsilon}^{2} \Phi + \dots
	\ .
	\label{defsuperpartner}
\end{align}
In the present case, it is convenient to deal with an expansion in powers of bilinears of $\epsilon$. This is denoted by the superscript $\left[ n \right]$, counting the number of bilinears.
Due to our choice of the background fields, we have
\begin{align}
	B^{\left[ n \right]}&= \frac{1}{2n!}\delta_{\epsilon}^{2 n}B
	\ ,&
	F^{\left[ n \right]}&= \frac{1}{(2n-1)!}\delta_{\epsilon}^{2 n - 1}F
	\ ,&
	n>0 \ ,
	\label{defTransf}
\end{align}
where $B$ and $F$ are generic bosonic and fermionic fields respectively. Then, for fermionic fields $\left[ n \right]$ counts $n-1$ bilinears plus a spinor $\epsilon$ while for bosonic fields it indicates $n$ bilinears. The $n=0$ case represents the background fields
\begin{align}
	e^{\left[ 0 \right]}{}_{M}^{A} & =  \left. e_{M}^{A}\right|_{BTZ} \ ,
	& \psi^{[0]}{}_{M} = 0 \ ,&
	& A^{[0]}{}_{M} = 0 \ .&
	\label{back1}
\end{align}
In the following we define the following real bilinears
\begin{align}
	\bB_{0} = &
	- i \zeta^{\dagger}\zeta
	\ ,
	& \bB_{1} = &
	- i \zeta^{\dagger}\sigma_{1}\zeta
	\ ,	
	& \bB_{2} = &
	i\zeta^{\dagger}\sigma_{2}\zeta
	\ ,
	& \bB_{3} =  &
	- i \zeta^{\dagger}\sigma_{3}\zeta
	\ ,
	\label{realBil0}
\end{align}
due to the anticommutative nature of $\zeta_{1}$ and $\zeta_{2}$, these identities hold
\begin{align}
	&\bB_{1}^{2}=\bB_{2}^{2}=\bB_{3}^{2}=-\bB_{0}^{2}\ ,&
	&\bB_{i}\bB_{j}=0\,,\, i\neq j\ ,&
	&\bB_{0}^{n}=0\,,\, n>2
	\ .&
	\label{fierz1}
\end{align}
In the following, $M,J$ are defined as in (\ref{boost1}) and we replace $\zeta \rightarrow \boldsymbol{\zeta}$ to highlight the fermionic contributions. The gravitino reads
\begin{align}
	\psi^{\left[ 1 \right]}
	= &
	\frac{1}{8 r \sqrt{r}}
	\left[
		\sigma_{1} \left[ - J \left( 1 + r \right) - 2 r \left( r - 1 \right)\left( r - N \right) \right]
\acI	\quad	
		+
		i\sigma_{2} \left[ - J \left( r - 1 \right) - 2 r \left( r + 1 \right)\left( r - N \right) \right]
\acI	\quad
		+
		\left( \sigma_{0} + \sigma_{3} \right) r \left( t - x \right)\left( - J - 2 r^{2} + 2 r N \right)
		\right] \boldsymbol{\zeta} \ \left( \dd t - \dd x \right)
\ac
		+
	\frac{1}{ 8 r^{2} \sqrt{r} N }
	\left( J - 2 r^{2} + 2 r N \right)
	\left[
		\sigma_{0} \left( r - 1 \right)
		-
		\sigma_{3} \left( r + 1 \right)
		+
		\left( \sigma_{1} - i \sigma_{2} \right) r \left( t - x \right)
		\right] \boldsymbol{\zeta} \ \dd r
	\ ,
	\label{grav}
\end{align}
and
\begin{align}
	\psi^{\left[ 2 \right]}
	= &
	\frac{1}{192 r^{2} \sqrt{r}}
	\left[
	i \bB_{3}
		\left(
			\left[ 1 - r^{2} \left( -1 + (t-x)^{2} \right) \right]
			-
			2 r( r - N)
			\left[ 1 + r^{2} \left( -1 + (t-x)^{2} \right) \right]
		 \right)
		 \times
\accI		
		 \qquad\times
		 \left(
		 	\sigma_{3} (1-r)
		 	+
		 	\sigma_{0} (1+r)
		 	+
		 	(\sigma_{1}-i \sigma_{2}) r (t-x)
		 \right)
\acI
	 +
	 i \bB_{0}
		 \left(
			\left[ 1 - r^{2} \left( 1 + (t-x)^{2} \right) \right]
			-
			2 r( r - N)
			\left[ 1 + r^{2} \left( 1 + (t-x)^{2} \right) \right]
		 \right)
		 \times
\accI
		 \qquad\times
		 \left(
		 	\sigma_{3} (1-r)
		 	+
		 	\sigma_{0} (1+r)
		 	+
		 	(\sigma_{1}-i \sigma_{2}) r (t-x)
		 \right)
\acI
	-
	2 \bB_{2} r
		\left(
			\sigma_{1}
				\left[
				J (1+r)
				+
				2 r (r-1) (r-N)
				\right]
			+
			i \sigma_{2}
				\left[
				J (r-1)
				+
				2 r (r+1) (r-N)
				\right]
\acII
			\qquad+
			(\sigma_{0}+\sigma_{3})
			r
			\left( J + 2 r^{2} - 2 r N \right)(t - x)
		\right)
\acI
	+
	2 i \bB_{1} r^{2}
	\left( - J - 2 r^{2} + 2 r N \right)
		\left(
		 	\sigma_{3} (1-r)
		 	+
		 	\sigma_{0} (1+r)
		 	+
		 	(\sigma_{1}-i \sigma_{2}) r (t-x)
		 \right)
	\right] \,\boldsymbol{\zeta}\, \left( \dd t - \dd x \right)
\ac	
	+
	\frac{J - 2 r^{2} + 2 r N}{96 r^{2} N \sqrt{r}}
	\left[
		i \left( - \bB_{1} + (\bB_{0} - \bB_{3}) (t-x) \right)
		\times
\accI
		\qquad\times
		\left(
		\sigma_{3} (1-r)
		+
		\sigma_{0} (1+r)
		+
		\left( \sigma_{1} -i \sigma_{2} \right) r (t-x)
		\right)
\acI
		+
		\bB_{2}
		\left(
		\sigma_{0}(r-1)
		-
		\sigma_{3}(1+r)
		+
		\left( \sigma_{1}-i\sigma_{2} \right)
		r
		(t-x
		\right)		
		\right]
		\,\boldsymbol{\zeta}\,\dd r
		\ .
        \label{grav2}
\end{align}
The metric corrections are
\begin{align}
	g^{\left[ 1 \right]}
	& =
	\frac{1}{4}
	\left(
	M - r^{2} + r N
	\right) \bB_{2} \dd t^{2}
	-
	\frac{1}{8}
	\left(
	J + 2 M -2 r^{2} + 2 r N
	\right) \bB_{2} \dd t \dd x
+ \nonumber \\ &
	+
	\frac{1}{8} J \bB_{2} \dd x^{2}
	-
	\frac{1}{8 r^{2} N^{2}}
	\left(
	J - 2 r^{2} + 2 r N
	\right) \bB_{2} \dd r^{2}
	\ ,
	\label{wig1}
\end{align}
and
\begin{align}
	g^{\left[ 2 \right]}
	& =
	\frac{1}{192}
	\left( 7M -10 r^{2} + 10 r N \right) \bB_{2}^{2} \dd t^{2}
	+
	\frac{1}{192}
	\left(
	2 J + 3 M - 6 r^{2} + 6 r N
	\right) \bB_{2}^{2} \dd x^{2}
+ \nonumber \\ &
	-
	\frac{1}{96}
	\left(
	J + 5 M - 8 r^{2} + 8 r N
	\right) \bB_{2}^{2} \dd t \dd x
+ \nonumber \\ &
	+
	\frac{1}{384 r^{4} N^{2}}
	\left[
		3 J^{2} - 6 r^{2} M
		+ 20 r^{3} \left( r - N \right)
		- 2 J r \left( 5 r -3 N \right)
		\right] \bB_{2}^{2} \dd r^{2}
	\ .
	\label{wig2}
\end{align}
The gauge field one--form is
\begin{align}
	A^{\left[ 1 \right]} = &
	\frac{1}{32 r^{2}}
	\left[
		\left( J - 2 r^{2} + 2 r N \right) \left( \bB_{3} + \bB_{0} \right)
		+
		r^{2} \left( J + 2 r^{2} - 2 r N \right)
		\left(
		\left( 1 - r^{2}\left( t-x \right)^{2} \right) \bB_{3}
\acII
		-
		\left( 1 + r^{2} \left( t - x \right)^{2} \right) \bB_{0}
		-
		2 \left( t - x \right) \bB_{1}
		\right)
	\right] \left( \dd t - \dd x \right)
\ac
	-
	\frac{1}{16 r N}
	\left( J - 2 r^{2} + 2 r N \right)
	\left( \bB_{1} + \left( \bB_{0} + \bB_{3} \right) \left( t-x \right) \right)
	\dd r
	\ ,
	\label{deltaGauge}
\end{align}
at second order, the gauge field is zero. Notice that in the large $r$ expansion $A^{\left[ 1 \right]}_{r} = O\left( \frac{1}{r^{3}} \right)$. As expected, the fermionic corrections collapse in the $AdS_{3}$ limit $M \rightarrow 0$, $J \rightarrow 0$. Note that the metric correction (wig) does not depend upon the boundary coordinates $x$,$t$.
Moreover, there is no off--diagonal corrections in the $rt$ and $rx$ components.
Last remark: Notice that the metric does not depend on boundary coordinates $t$ and $x$, that is the two translational isometries of BTZ black hole are preserved by the wig. This allows to define the wig's mass and the angular momentum.


\subsection{Large \boldmath{$r$} Results}

Here we present the obtained results in large $r$ expansion.
To simplify the notation, we define the following expressions
\begin{align}
	\bF & =
	\left[ 1 + ( t - x )^{2} \right] \bB_{0}
	+
	\left[ - 1 + ( t - x )^{2} \right] \bB_{3}
	+
	2 ( t - x ) \bB_{1}
	\ , \quad\quad \bF^2 = 0\,.
	\label{LargeR1}
\end{align}
and
\begin{align}
	\bH = \frac{1}{8} \bB_{2} + \frac{1}{96} \bB^{2}_{2}
	\ .
	\label{LargeR1a}
\end{align}
The gravitino reads
\begin{align}
	\psi & \sim
	\frac{J+M}{192}
	\left[
		\sqrt{r}\, \bF
		\left(
		i \sigma_{0} + i\sigma_{3}
		-
		( i\sigma_{1} + \sigma_{2} ) ( t - x )
		\right)
\acI	
		-
		\frac{1}{\sqrt{r}}
		\left(
		2 ( 12 + \bB_{2})\left( \sigma_{1} + i \sigma_{2} \right)
		+
		\left(
		2 ( 12 + \bB_{2} ) ( t - x ) + i \bF
		\right)
		\left( \sigma_{0} + \sigma_{3} \right)
		\right)
	 \right] \boldsymbol{\zeta} \left( \dd t - \dd x \right)
\ac	
	 +
	 \frac{J-M}{96 r^{2} \sqrt{r}}
	 \left[
	 12 + \bB_{2} - i \bB_{1}
	 -
	 i \left( \bB_{0} + \bB_{3} \right)( t - x )		
	\right]
	\left[ \sigma_{0} - \sigma_{3}
	+
	(\sigma_{1} - i \sigma_{2})( t - x )
	\right]\boldsymbol{\zeta} \, \dd r
	\ .
	\label{LargeRgrav}
\end{align}
The full metric at large $r$ is
\begin{align}
	g \sim &
	-
	\left[ r^{2} - M \left( 1 + \bH  \right) \right] \dd t^{2}
	-
	\left[ J + (M+J)\bH  \right] \dd t \dd x
	\ac
	+
	\left[ r^2 + J \bH \right]  \dd x^{2}
	+
	\frac{1}{r^{2}}
	\left[ 1+ \frac{1}{r^{2}} \left( M -(M-J)\bH  \right) \right]\dd r^{2}
	\ ,
	\label{glarger}
\end{align}
that is
\begin{align}
	g \sim &
	-(r^{2} -M) \dd t^{2}
	- J \dd t \dd x
	+ r^{2} \dd x^{2}
	+\frac{1}{r^{2}}  \left( 1+\frac{M}{r^{2}} \right) \dd r^{2}
\ac
	+
	\bH
	\left[
		M \dd t^{2} - (M+J) \dd t \dd x + J \dd x^{2} - \frac{1}{r^{4}} (M-J) \dd r^{2}
		\right]
		\ .
	\label{glarger1}
\end{align}
Last, the large $r$ gauge field is
\begin{align}
	A & \sim
	-\frac{J+M}{32}\, \bF (\dd t - \dd x)
	-
	\frac{J- M}{16 r^{3}}
	\left[ \bB_{1} + ( \bB_{0} + \bB_{3} )( t - x ) \right] \dd r
	\ .
	\label{LagerRgauge}
\end{align}
In this limit we can rewrite the vielbein and the spin connection for the metric (\ref{glarger}). They read
\begin{align}
	e^0 & =
	\left( r - \frac{M}{2 r} \right) \dd t +
	\frac{M}{2 r} 	\bH (\dd x - \dd t)
\ , \nonumber \\
	e^1 & =
	\left(
		\frac{1}{r} +
		\frac{M}{2 r^{3}}
		+ \frac{M-J}{2 r^{3}} \bH
	\right)\dd r	
\ ,\nonumber  \\
	e^2 & =
	\left( r - \frac{J}{2 r} \right) \dd x
	+
	\frac{J}{2 r} \bH (\dd x - \dd t)
\ ,
\end{align}
and
\begin{align}
	\omega^{01}
	& = \left(r \dd t -\frac{J}{2 r} \dd x \right) + \bH \frac{J}{2 r} \left(\dd t - \dd x\right)
	\ ,\nonumber\\
	\omega^{02}
	& =
	- \frac{1}{2 r^{3}}
	\left[
	J + ( J - M ) \bH
	\right] \dd r
\ ,\nonumber \\
	\omega^{12}
	& =
	- \left(r - \frac{M}{2 r}\right) \dd x - \bH \frac{M}{2 r}\left( \dd t - \dd x\right)
\ .
\end{align}
The large--$r$ curvature $2$--form is computed from the definition in (\ref{action0}).
The non--zero components are
\begin{align}
	R^{01}
	& =
	\frac{M}{2 r^{2}} \bH \dd r \wedge \dd x
	+ \left( 1 - \frac{J}{2 r^{2}}  \bH \right) \dd r \wedge \dd t
\ , \nonumber \\
	R^{02}
	& =
	\left[
		r^{2} + \frac{J}{2} \bH - \frac{M}{2}\left( 1 + \bH \right)
	\right] \dd x \wedge \dd t
\ , \nonumber \\
	R^{12}
	& =
	\frac{J}{2 r^{2}} \left( 1 + \bH \right) \dd r \wedge \dd t
	-
	\left[
		1 + \frac{M}{2 r^{2}} \left( 1 + \bH \right)
	\right] \dd r \wedge \dd x
	\ .
	\label{LargeRcurvature}
\end{align}
It is easy to show that the equations of motion (\ref{EoMs1}) are satisfied.
In particular, in the large $r$ limit, the term $- \frac{\Lambda}{4} \varepsilon^{AB}{}_{C} \bar\psi\Gamma^{C}\psi$ is subleading order, hence it does not contribute to the equations of motion.


\section{Linearized Boundary Equations}

We refer to \cite{Bhattacharyya:2008jc} to compute the Navier-Stokes equations dual to Einstein's equations, for a boosted and dilatate BTZ.
However our method is slightly different: our fermionic degrees of freedom induce a non--zero torsion that must be taken into account to
verify Einstein's equations. Moreover, we derive a new set of equations of motion which emerges from the gravitino field equation.

Technically for computing the Riemann tensor we use the spin connection formalism:
\begin{equation}
R^{ab} = \dd \omega^{ab} + \omega^{a}_{\phantom{a}c} \wedge \omega^{cb}
\ .
\end{equation}
In the form language it is easy to check that -- working at first order and expanding $\hat b$ around $1$ (no dilatation) and $w$ around $0$ (no boost) -- the boosted metric together with the boosted wig, satisfies (\ref{EoMs1}).

As explained in \cite{Bhattacharyya:2008jc} when we promote the parameters to local functions of the boundary coordinates the obtained metric is not a solution of the equations of motion anymore.
In order to reconstruct a solution, we must constrain the parameters to fulfill some equations which represent the equations of motion for the boundary fluid and we also need to add corrections to the metric.
Consequently, also the parameters must be modified accordingly.\footnote{The interested reader shall refer to \cite{Bhattacharyya:2008jc,Rangamani:2009xk} for further details.}
Since we work in a perturbative procedure, the metric is corrected order by order in the derivative expansion:
\begin{align}
	g \rightarrow g^{(0)} + g^{(1)} + \dots
	\ ,
	\label{gcorrection}
\end{align}
where $g^{(0)}$ represents the deformed metric and $g^{(i)}$ for $i>0$ are the metric corrections at the order $i$ in boundary derivatives.
In the following we limit our discussion at first order, namely we consider only $g^{(1)}$ correction.
Imposing the equations of motion on $g$, two kinds of equations are found.
The first one comprehends equations involving only derivatives of local parameters and no terms belonging to the metric correction $g^{(1)}$: these are the linearized Navier-Stokes equations for local parameters in $2$--dimensions (the conformal factor in front of the ``divergence'' of $w$ is just $1$).
The second set of equations generically could depend on both parameters and $g^{(1)}$ components.
These are called dynamical equations and are used to obtain the metric correction $g^{(1)}$ in terms of the derivatives of the parameters.
Notice that being $g^{(1)}$ a  first--order contribution, it will depend on a single derivative of the parameters.
In general, it is convenient to classify them according to the representations of the little group $SO(1,d-1)$.

As a warming--up exercise, we compute the NS equations derived from the metric variation due to the $AdS_3$ isometries acting on the global BTZ black hole metric
\begin{equation}
\delta g = \mathcal{L}_{\xi}\left(g_{BH}\right)
\ ,
\end{equation}
where $g_{BH}$ is the BTZ metric (\ref{BTZ1}) and $\xi$ are defined in (\ref{KVglobal}). We observe that all isometries are broken, except the ones generated by $e_0$ and $f_0$.

We now proceed as follows. First of all, we promote all Killing vectors parameters to local functions of the boundary coordinates ($t$ and $\phi$); then we check Einstein's equations for the metric
\begin{equation}
	g = g_{BH} + \delta g + g^{(1)}
\end{equation}
which, as expected, are not satisfied. Yet, imposing them yields the following equations for the functions $b_0,d_0\ldots$ expanding near $t=\phi =0$ we get:
\begin{align}
& J_0 \left[\partial_{\phi} \left(b_0 + d_0\right) + \partial_t \left(b_0 - d_0\right)\right] - 2 \left(1 + M_0 \right) \partial_t \left(b_0 + d_0\right) = 0 \ , \nonumber \\
& J_0 \left[\partial_{t} \left(b_0 + d_0\right) + \partial_{\phi} \left(b_0 - d_0\right)\right] - 2 \left(1 + M_0 \right) \partial_{\phi} \left(b_0 + d_0\right) = 0 \ .
\label{NS1}
\end{align}
Note that these equations are computed in the global $AdS_3$; for other choices of neighborhoods, for example $t= \phi = \pi/2$, similar equations for the other parameters are obtained.
These are the Navier-Stokes equations derived by the global metric. As expected in the empty $AdS_3$ limit $J_0 \rightarrow 0, M_0 \rightarrow -1$ they are satisfied identically.

For what concerns the dynamical equations, the $3$--dimensional case is slightly different from higher dimensional cases. In fact, once the constraint equations are satisfied, no further corrections are needed and Einstein's equations are satisfied up to the first order in the derivative expansion.
Therefore $g^{(1)}$ can be set to zero. This is an important result since it implies that we are dealing with a perfect fluid with no dissipative corrections (contrary to \cite{Bhattacharyya:2008jc}, where the non-vanishing first order corrections corresponded to the shear tensor) and with second order, non--dissipative transport coefficients.

%

\subsection{Corrected NS Equations}

Having added fermionic fields to our scheme, the Navier-Stokes equations are now dual to the equations of motion derived from the $\mathcal{N}=2$, $D=3$ $AdS-$supergravity action (\ref{action1}).\footnote{Note that $\mathcal{N}=2$ supergravity Killing spinors do not suffer the problem pointed out by Gibbons in \cite{Gibbons:1981ux}. In fact, their behavior is stable even in the large $r$ limit, in contrast with in $\mathcal{N}=1$ theories.}

Once the fermionic bilinears are taken into account, imposing equations of motion (\ref{EoMs1}) and taking the large $r$ limit, we find:
\begin{align}\label{NSwig1}
&M_0\left[\partial_x b + \partial_t w - \frac{1}{16}\left(\partial_x +\partial_t\right)\bB_2\right]=0\ ,
\nonumber\\
&M_0\left[\partial_t  b + \partial_x w - \frac{1}{16}\left(\partial_x  +\partial_t\right)\bB_2\right]=0\ .
\end{align}
These are the Navier-Stokes equations for the Poincar\'e patch (cfr. (\ref{NS1})). Note that in this case they are identically satisfied if $M_0$ is set to zero.

Remarkably, as in the case of BTZ in global coordinates without fermionic wig, all the equations of motion lead to (\ref{NSwig1}).
Therefore the first order metric correction $g^{(1)}$ can be set to zero.
As in the previous section, this means that the conformal fluid on the boundary have non -- dissipative first order corrections, as expected for a two dimensional  conformal fluid.

\subsection{Dirac-type equation}

This is a truly original study, since nobody takes the deformation of Rarita-Schwinger equation in to account in the present framework. Therefore we explain carefully the technique adopted.

We proceed as follows: first we consider the solution of $\mathcal{D} \psi = 0$  where the spinor $\zetab$ is a constant field (zero mode) and we promote it to be local upon boundary coordinates. This implies that we can rewrite the gravitino field proportional to the fermionic field itself:
\begin{align}
	\psi_{M} = \Upsilon_{M} \zetab
	\ ,
	\label{RS1}
\end{align}
where $\Upsilon_{M}$ is a generic $2\times2$ matrix which depends on the coordinates $t,r,x$ (and in principles also on the bilinears) that can be decomposed on the basis of the Pauli matrices (\ref{gammaBTZ}) and the identity.
Notice that since $\psi_t = - \psi_x$ we have
\begin{align}
	\Upsilon_{x} = - \Upsilon_{t}
	\ .
	\label{RS2}
\end{align}
Consequently, the equations of motion read
\begin{align}
	\varepsilon^{MNR} {\cal D}_{N} \left(\Upsilon_{R} \zetab \right) = 0
	\ ,
	\label{EoMgrav}
\end{align}
By promoting $\zetab$ to be local on the boundary coordinates $t,x$ and using the equations of motion for the constant $\zetab$,
eqs.~(\ref{EoMgrav}) become
\begin{align}
	\varepsilon^{MNR} \Upsilon_{R} \partial_{N} \zetab (t,x) = 0
	\ .
	\label{EoMgrav2}
\end{align}
Being $\partial_{N} \zetab (t,x)$ a spinor, it can be written as a linear transformation of the spinor $\zetab(t,x)$ itself
\begin{align}
	\partial_{N} \zetab (t,x) = \Theta_{N} \zetab (t,x)
	\ ,
	\label{RS3}
\end{align}
where $\Theta_{N}$ is a $2\times2$ matrix. Notice that since $\zetab$ is not a function of the radial coordinate $r$, we have
\begin{align}
	&\Theta_{R} = \left\{\Theta_t (t,x),0,\Theta_x (t,x) \right\}
	\ .
	\label{RS5}
\end{align}
Eqs.~(\ref{EoMgrav2}) then reduce to
\begin{align}
	\varepsilon^{MNR} \Upsilon_{R} \Theta_{N} \zetab (t,x) = 0
	\ ,
	\label{EoMgrav3}
\end{align}
which in components read
\begin{align}
	\left( \Upsilon_{r} \Theta_{x} - \Upsilon_{x} \Theta_{r}  \right) \zetab = 0
	\ , \qquad
	\left( \Upsilon_{t} \Theta_{x} - \Upsilon_{x} \Theta_{t} \right) \zetab = 0
	\ , \qquad
	\left( \Upsilon_{r} \Theta_{t} - \Upsilon_{t} \Theta_{r} \right) \zetab = 0
	\ .
	\label{RS4}
\end{align}
Using (\ref{RS2}) and (\ref{RS5}) we have
\begin{align}
	&\Theta_{x} = - \Theta_{t}\ ,&
	\Upsilon_{r} \Theta_{t} \zetab = 0 \ .
	\label{RS6}
\end{align}
Thus, there is only one independent  matrix $\Theta$:
\begin{align}
	\Theta_{t} \equiv
	\Theta =
	\left(
	\begin{array}{cc}
		\theta_{11} & \theta_{12} \\
		\theta_{21} & \theta_{22} \\
	\end{array}
	\right)
	\ .
	\label{RStheta}
\end{align}
Considering only the first order gravitino (\ref{grav}), after a straightforward computation  at leading order in $r\rightarrow\infty$ expansion we get
\begin{align}
	\Upsilon_{r} \sim
	\frac{1}{4} (J - M)
	r^{-5/2}
	\left(
	\begin{array}{cc}
	- 1/r & 0 \\
	(t - x) & 1 \\
	\end{array}
	\right)
	\ ,
	\label{RS7}
\end{align}
In $r\rightarrow\infty$ asymptotic limit the matrix $\Upsilon_{r}$ is no longer invertible, therefore the second equation of (\ref{RS6}) in that limit becomes:
\begin{align}
	\left[
	\theta_{21} + \theta_{11} (t - x)
	\right] \zeta_{1}	
	+
	\left[
	\theta_{22} + \theta_{12} (t - x)
	\right] \zeta_{2}
	=
	0
	\ ,
	\label{RS8}
\end{align}
where $\zeta_{1}$ and $\zeta_{2}$ are the Grassmann components of $\zetab$.
Solving for generic $\zeta_{1},\zeta_{2}$, we obtain
\begin{align}
	&\theta_{21}  = - (t-x) \theta_{11} \ ,&
	&\theta_{22}  = - (t-x) \theta_{12} \ .&
	\label{RS9}
\end{align}
Summing up the results, eqs.~(\ref{RS3}) read
\begin{align}
	& \partial_{t} \zeta_{1} = \theta_{11} \zeta_{1} + \theta_{12} \zeta_2 \ ,
	& \partial_{t} \zeta_{2} = -( t - x ) \left( \theta_{11} \zeta_{1} + \theta_{12} \zeta_2  \right)\ ,&
	\nonumber \\
	& \partial_{x} \zeta_{1} = - \theta_{11} \zeta_{1} - \theta_{12} \zeta_2 \ ,
	& \partial_{x} \zeta_{2} = + ( t - x ) \left( \theta_{11} \zeta_{1} + \theta_{12} \zeta_2  \right)\ ,&
	\label{RS10}
\end{align}
Notice that this implies
\begin{align}
	\left( \partial_{t} + \partial_{x}  \right) \zetab = 0\ .
	\label{derBil0}
\end{align}
From the definitions (\ref{realBil0}), we compute the bilinears derivatives
\begin{align}
	\partial_{t} \bB_{0}
	& =
	\bB_{0} \left[ \textrm{Re}\theta_{11} - ( t- x ) \textrm{Re}\theta_{12} \right]
	+
	\bB_{1} \left[ \textrm{Re}\theta_{12} - ( t- x ) \textrm{Re}\theta_{11} \right]
	+
	\nonumber\\
	&
	+
	\bB_{2} \left[ \textrm{Im}\theta_{12} + ( t- x ) \textrm{Im}\theta_{11} \right]
	+
	\bB_{3} \left[ \textrm{Re}\theta_{11} + ( t- x ) \textrm{Re}\theta_{12} \right]
	\ ,
	\label{derBil1}
\end{align}
\begin{align}
	\partial_{t} \bB_{1}
	& =
	\bB_{0} \left[ \textrm{Re}\theta_{12} - ( t- x ) \textrm{Re}\theta_{11} \right]
	+
	\bB_{1} \left[ \textrm{Re}\theta_{11} - ( t- x ) \textrm{Re}\theta_{12} \right]
	+
	\nonumber\\
	&
	-
	\bB_{2} \left[ \textrm{Im}\theta_{11} + ( t- x ) \textrm{Im}\theta_{12} \right]
	-
	\bB_{3} \left[ \textrm{Re}\theta_{12} + ( t- x ) \textrm{Re}\theta_{11} \right]
	\ ,
	\label{derBil2}
\end{align}
\begin{align}
	\partial_{t} \bB_{2}
	& =
	\bB_{0} \left[ \textrm{Im}\theta_{12} + ( t- x ) \textrm{Im}\theta_{11} \right]
	+
	\bB_{1} \left[ \textrm{Im}\theta_{11} + ( t- x ) \textrm{Im}\theta_{12} \right]
	+
	\nonumber\\
	&
	+
	\bB_{2} \left[ \textrm{Re}\theta_{11} - ( t- x ) \textrm{Re}\theta_{12} \right]
	-
	\bB_{3} \left[ \textrm{Im}\theta_{12} - ( t- x ) \textrm{Im}\theta_{11} \right]
	\ ,
	\label{derBil3}
\end{align}
\begin{align}
	\partial_{t} \bB_{3}
	& =
	\bB_{0} \left[ \textrm{Re}\theta_{11} + ( t- x ) \textrm{Re}\theta_{12} \right]
	+
	\bB_{1} \left[ \textrm{Re}\theta_{12} + ( t- x ) \textrm{Re}\theta_{11} \right]
	+
	\nonumber\\
	&
	+
	\bB_{2} \left[ \textrm{Im}\theta_{12} - ( t- x ) \textrm{Im}\theta_{11} \right]
	+
	\bB_{3} \left[ \textrm{Re}\theta_{11} - ( t- x ) \textrm{Re}\theta_{12} \right]
	\ .
	\label{derBil4}
\end{align}
where
\begin{align}
	& \textrm{Re}\theta = \frac{1}{2} ( \theta + \theta^{*} ) \ ,
	&
	 \textrm{Im}\theta = \frac{1}{2 i} ( \theta - \theta^{*} )
	\ .&
	\label{derBil31}
\end{align}
The $x$--derivative of bilinears satisfies
\begin{align}
	\partial_{x} \bB_{i} = - \partial_{t} \bB_{i} \ .
	\label{derBilx}
\end{align}
The last equation has a strong implication on the linearized Navier-Stokes equations (\ref{NSwig1}), indeed this implies
that the last term there vanishes. Therefore, the two sets of equations are decoupled at the linearized level. This yields the possibility of a
clear separation of the bosonic and fermionic degrees of freedom. It would be very interesting to study the complete non-linearized version
of these equations.


\section{Physics at the Horizon and at the Boundary}

\subsection{Energy--Momentum Tensor dual to BTZ black hole}
Using \cite{Balasubramanian:1999re} we compute the boundary energy--momentum tensor $T_0^{\mu\nu}$ for the boosted metric. 
Notice that Greek indices labels the boundary coordinates $t,x$.
Defining the normal vector $n^{M}$ to constant $r-$slice we can compute the extrinsic curvature
\begin{equation}
\kappa^{MN} = \frac{1}{2}\left(\nabla^{M} n^N - \nabla^{N} n^M \right)
\ ,
\end{equation}
and then
\begin{equation}
T^{MN} = \kappa^{MN} - (\kappa+1) \gamma^{MN}
\ ,
\end{equation}
where $\kappa$ is the trace of $\kappa^{MN}$ and $\gamma_{MN}$ is the boundary metric.
This turns out to be
\begin{align}
	T_0^{\mu\nu}
	=
	\frac{1}{2}\left(
	\begin{array}{cc}
	M & - J \\
	- J & M
	\end{array}
	\right)\ .
	\label{Tmunu1}
\end{align}
In order to get the usual form of perfect fluid energy--momentum tensor
\begin{align}
	T_0^{\mu\nu} = \eta^{\mu\nu} + 2 u^{\mu} u^{\nu}
	\ ,
	\label{Tmunu2}
\end{align}
it is sufficient to consider the case $J_{0}=0$.
Indeed the metric will acquire angular momentum due to the Lorentz transformation
as shown in (\ref{boost2}).
The fluid boundary energy--momentum tensor dual to the metric (\ref{BTZ1}) with $J_{0}=0, M_{0}\neq 0$
is the standard one for the perfect fluid in the rest frame.

Then we perform the boost transformation which switches on an angular momentum and modifies the mass parameter
\begin{align}
	&
	M =  \frac{1 + w^{2}}{1 - w^{2}} M_{0}
	\ ,
	&
	J =   \frac{2 w}{1 - w^{2}} M_{0}
	\ .
	&
	\label{boostNoJ}
\end{align}
Notice that our results are in perfect agreement with \cite{Martinez:1999qi} since we obtain the extremality condition
once we set $\left|w \right|=1$.
Starting from the boosted metric, \textit{i.e.} the metric (\ref{BTZ1}) in which $M_0$ and $J_0$ has been replaced with eqs. (\ref{boost2}) and $r$ with (\ref{boost3}), the computation of $T_0^{\mu\nu}$ yields
\begin{align}
	T_0^{\mu\nu}
	=
	M_0 \gamma \left(
	\begin{array}{cc}
	1  & \beta \\
	\beta & 1
	\end{array}
	\right)\ .
	\label{Tmunu3}
\end{align}
where $\gamma$ and $\beta$ are defined in (\ref{boost4}).
Setting
\begin{align}
	&
	u^{0} = \frac{1}{\sqrt{1 - w^{2}}}
	\ ,
	&
	u^{1} = -\frac{w}{ \sqrt{1 - w^{2}}}
	\ ,
	&
	\label{Tmunu4}
\end{align}
we find precisely (\ref{Tmunu2}) where $u^{\mu}$ is the normalized fluid velocity ({\it i.e.} $u^{\mu}u_{\mu}=-1$).

It is now straightforward to recover the variation of $T_0^{\mu\nu}$ due to a dilatation. In fact, being proportional to $M_0$, it scales as
\begin{equation}
	T_0^{\mu\nu} \rightarrow T^{\mu\nu} = \frac{1}{\hat{b}^2}T_0^{\mu\nu}
\ .
\end{equation}

Using the results obtained in \cite{Gentile:2012tu} we compute the Brown-York energy--momentum tensor dual to the BTZ black hole with fermionic wig. Note that this is an exact result since the series in the fermionic bilinears naturally truncates at second order:
\begin{align}
\label{TmunuNew0}
T^{\mu\nu} &  = \frac{M_0}{2 \hat{b}^2} \left(1+\bH\right) \left(\eta^{\mu\nu} + 2 u^{\mu} u^{\nu}\right) - \frac{M_0}{2 \hat{b}^2} ~ \bH ~ \varepsilon^{\mu\sigma}\left(\delta^{\nu}_{\phantom{\nu}\sigma} + 2 u^{\nu} u_{\sigma}\right) \ ,
\end{align}

Eq.~(\ref{TmunuNew0}) can be recast in the following form
\begin{align}
T^{\mu\nu} &  = \frac{M_0}{2 \hat{b}^2} \left(1+\bH\right) \left(\eta^{\mu\nu} + 2 u^{\mu} u^{\nu}\right) - \frac{M_0}{\hat{b}^2} ~\bH~ \varepsilon^{\left(\mu|\sigma\right.} u^{\left.\nu\right)} u_{\sigma} \ .
\label{TmunuNew1}
\end{align}
By assuming that the bilinears contained in $\bH$ are local quantities, the equations for the conservation of the energy-momentum tensor $T^{\mu\nu}$ lead to differential
equations involving also the bilinears.
At linearized level these equations reduce to eqs.~(\ref{NSwig1}).

\subsection{Redefining the Velocity}

At first glance, equation (\ref{TmunuNew1}) reveals a parity-violating term. This term has been
studied in \cite{Dubovsky:2011sk}, where anomalous fluid are considered, and they concluded that the most general form for it is
\begin{equation}
\label{deltaT}
\Delta T^{\mu\nu} = - \left[\mu^2 C + \alpha \left(T^2 + \frac{2 n T \mu}{s}\right)\right] u^{\left( \mu \right.} \tilde{u}^{\left. \nu\right)}
\ ,
\end{equation}
where $\tilde{u}^{\mu} = \varepsilon^{\mu\nu} u_{\nu}$, $C$ is the coefficient of the anomaly, $T$ is the temperature, $n$ is the fluid charge density, $s$ the entropy density, $\mu$ is the chemical potential and $\alpha$ an arbitrary integration constant.

Nevertheless, as pointed out by \cite{Jain:2012rh}, the anomaly require the following background metric and gauge field
\begin{align}
\dd s^2 &= -e^{2 \sigma} \left(\dd t + a_1 \dd x\right)^2 + g_{11} \dd x^2 \ , \\ \nonumber
A &= A_0 \dd t + A_1 \dd x \ .
\label{metric1plus1}
\end{align}
where $\sigma, a_1$ and $g_{11}$ are functions of $x,t$.
In the present case we have
\begin{align}
\dd s^2 =& -\dd t^2 +  \dd x^2 \ ,\\
A =& -\frac{1}{32}\left(M+ J\right) \left[2 \bB_1 \left(t - x\right) +
    \bB_3 \left(-1 + t^2 - 2 t x + x^2\right) \right. + \nonumber \\
    & \left. +\bB_0 \left(1 + t^2 - 2 t x + x^2\right)\right] \left(\dd t - \dd x\right) \ ,
\end{align}
and, comparing with (\ref{metric1plus1}) we get
\begin{align}
& \sigma = 0\ ,&
& a_1 = 0\ ,&
& g_{11} = 0\ , &
& F = \dd A = 0 \ .
\end{align}
Using the Poincar\'e lemma, we conclude that $A = \dd \lambda$ globally, therefore $A$ is a pure gauge and our theory is not anomalous.

Thus $C = 0$ leads to
\begin{equation}
\label{deltaTfinal}
\Delta T^{\mu\nu} = 2 \alpha T^2  u^{\left( \mu \right.} \tilde{u}^{\left. \nu\right)} \ .
\end{equation}

As explained in \cite{Dubovsky:2011sk}, in absence of an anomaly there is the freedom to add this term and it corresponds to a choice of the entropy current. In fact, it is possible to recast the energy--momentum tensor (\ref{TmunuNew1}) in the perfect fluid form
\begin{align}
	T^{\mu\nu}
	& =
	(1 + \frac{1}{8} \bB_{2} + \frac{1}{384} \bB_{2}^2) \frac{M_{0}}{2 \hat{b}^{2}}
	\left(
	2 U^{\mu} U^{\nu} + \eta^{\mu\nu}
	\right)
	\ ,
	\label{TmunuRecasted1}
\end{align}
through a redefinition of the velocity field
\begin{align}
	u^{\mu} \rightarrow U^{\mu}
	=
	\left( 1 + \frac{1}{512} \bB_{2}^{2} \right) u^{\mu}
	-
	\frac{1}{16} \left( \bB_{2} - \frac{1}{24} \bB_{2}^{2} \right) \tilde u^\mu \ .
	\label{NewU1}
\end{align}
Note that $U^{\mu}$ is correctly normalized to $-1$. Recalling the conformal thermodynamics identities \cite{Bhattacharyya:2008mz}
\begin{align}
& b = \frac{1}{2 \pi T} \ ,
&p = \rho = \frac{M_0}{2 b^2} = 2 \pi^2 T^2
\ ,
\end{align}
we immediately see that the temperature gets a shift due to the presence of bilinears
\begin{equation}
T' = T \left(1 + \frac{\langle\bB_2\rangle}{16} - \frac{\langle\bB_2^2\rangle}{1536}\right)\ ,
\end{equation}
where the brackets denotes the vev of the bilinears.

We have to make one important remark: the expression of the temperature in terms of the bilinear acquires a numerical value whenever the bilinear 
have a vev computed by path integral means (we have to recall that Grassmann numbers pertain only to the quantum realm). 
The procedure is similar to what is usually done in the case of solitons in gauge theories and supergravity 
\cite{Amati:1988ft,Konishi:1988mb} and the gravitinos condensate leads to non-vanishing vev of the bilinears interested in the previous formula. 
In the case of BTZ black hole, the gravitational action evaluated on the solution with the wig has never been computed and it will be presented elsewhere. 

\subsection{Horizon and Entropy}

In the following we present the entropy computed from the wig of the BTZ in global coordinates \cite{Gentile:2012tu}.
By direct computation we notice that the event horizon radius
\begin{align}
	r_{\pm}^{2}
	& =
	\frac{1}{2} \left( M \pm \sqrt{M^{2} - J^{2}}  \right) \ .	
	\label{horizon}
\end{align}
is not modified by the presence of the fermionic wig.
We can compute the entropy from Bekenstein--Hawking formula
\begin{align}
	S & = 	\frac{1}{4}A_{H} \ ,
	\label{BekensteinHawking}
\end{align}
where the area of the horizon reads
\begin{align}
	A_{H}
	=
	\int_{0}^{2\pi} \sqrt{\mathbf{g}_{\phi\phi}(r_{+})}
	\dd \phi \ ,
	\label{Area}
\end{align}
and is computed using the complete metric with the wig.
We obtain the following result
\begin{align}
	S & =
	\frac{\pi}{2}
	\left[
		r_{+}
		+
		\langle\bB_{2}\rangle \frac{J}{16 r_{+}}
		+
		\langle\bB_{2}^{2}\rangle \frac{1}{512 r^{3}_{+}}
		\left(
		J^{2} + 2 r^{2}_{+} \left( J - 2 - 2M \right)
		\right)
		\right]
	\ ,
	\label{Entropy}
\end{align}
where we take the vev for the bilinears. As can be seen the entropy of the black hole is modified by the presence of the wig confirming that
we are studying a new solution of the theory where the fermions play a fundamental r\^ole. Setting $J =0$ the
first order correction vanishes. This could also have been checked by a simple infinitesimal calculation. Nonetheless, the
second order corrections do not vanish. In particular for vanishing angular momentum the third term in the above equation
becomes proportional to $M+1$ which vanishes for $M=-1$, namely global anti-de Sitter.

By setting $J=M$ in the case of extremal solution, we find the simplified formula
\begin{align}
	S & =
	\frac{\pi}{2}  \sqrt{2 M}
	\left(
		\frac{1}{2}
		+
		\frac{1}{16} \langle\bB_{2}\rangle
		+
		\frac{M-2}{128 M} \langle\bB^{2}_{2}\rangle
		\right)
		\ ,
	\label{Entropy2}
\end{align}
showing that also in the case of extremal black hole the entropy is modified.


\subsection{Conserved Charges}

Here we compute the conserved charges associated with the isometries of the BTZ black hole.
We use holographic technique based on the boundary energy momentum tensor $T_{\mu\nu}$ \cite{Gentile:2011jt,Brown:1992br,Balasubramanian:1999re,Henneaux:1999ib,Henneaux:1984ei}.
To perform the computation we cast the boundary metric $\gamma_{\mu\nu}$ in ADM--like form
\begin{align}
	\gamma_{\mu\nu} \dd x^{\mu} \dd x^{\nu}
	& =
	- N_{\Sigma}^{2} \dd t^{2}
	+
	\sigma
	( 	\dd \phi + N^{\phi}_{\Sigma} \dd t )^{2}
	\ ,
	\label{ADM1}
\end{align}
where $\Sigma$ is the $2$--dimensional surface at constant time and the integration is over a circle at spacelike infinity.
The conserved charges associated to the Killing vectors $\xi$ are defined as
\begin{align}
	Q_{\xi} & =
	\lim_{r\rightarrow\infty}\int_{V} \dd x \sqrt{\sigma} u^{\mu} T_{\mu\nu} \xi^{\nu}
	\label{charges1}
\end{align}
where  $u^{\mu}=N_{\Sigma}^{-1}\delta^{\mu t}$ is the timelike unit vector normal to $\Sigma$.

In the present case, the wig does not depend on $t$ and $x$. Thus, the two resulting Killing vectors are
\begin{align}
	\xi_{1}^{\mu} &= \delta^{\mu t}
	\ ,
	&\xi_{2}^{\mu} = - \delta^{\mu x}
	\ .&
	\label{charges2}
\end{align}
The associated charges are respectively the mass $M_{tot}$ and the angular momentum $J_{tot}$. After a short computation we find
\begin{align}
	M_{tot} & =
	M
	+
	\frac{1}{8}\left( M + J \right) \langle \bH \rangle	
	\ , \nonumber \\
	J_{tot} & =
	J
	+
	\frac{1}{8}\left( M + J \right) \langle \bH \rangle
	\ ,
	\label{MandJ}
\end{align}
where $\bH$ is defined in (\ref{LargeR1a}).

\section{Conclusions}

In this paper, we analyze in detail the structure of the fermionic wig for the BTZ black hole. We derive the fermionic corrections to the mass and to the angular momentum
of the BTZ black hole. In addition, we compute the entropy of the black hole which also shows new terms depending on the vev's of the fermionic bilinears.
Finally, we also present the $r$-large expressions for the several geometrical quantities in the presence of the fermionic corrections.

On the other hand, by following the rules of the fluid/gravity correspondence, we derive the boundary equations of motion for a supersymmetric fluid. This means
a set of bosonic equations of motion, but also some Dirac-type equation for the supersymmetric long range d.o.f. of the fluid. The computations is performed at the
first order. Nonetheless, we are also able to provide the energy-momentum tensor which is cast in a form from which one can read the thermodynamic quantities.

\section*{Acknowledgements}

We thank S. Ferrara, A. Marrani, G. Policastro, M. Porrati, and  L. Sommovigo 
for useful discussions. This work was supported in part by the MIUR-PRIN contract 2009-KHZKRX.

\end{document}